\documentclass[10pt,a4paper,twocolumn,english,prl,showpacs,noshowkeys,nosuperscriptaddress]{revtex4}

\usepackage{mathpazo}

\usepackage[T1]{fontenc}
\usepackage[latin1]{inputenc}
\usepackage[dvips]{color}
\usepackage{babel}
\usepackage{amsmath}
\usepackage{amssymb}
\usepackage[dvips]{graphicx}
\usepackage{esint}
\usepackage[unicode=true,pdfusetitle,
 bookmarks=true,bookmarksnumbered=false,bookmarksopen=false,
 breaklinks=false,pdfborder={0 0 0},backref=false,colorlinks=true]
 {hyperref}
\usepackage{breakurl}

\makeatletter

\@ifundefined{textcolor}{}
{%
 \definecolor{BLACK}{gray}{0}
 \definecolor{WHITE}{gray}{1}
 \definecolor{RED}{rgb}{1,0,0}
 \definecolor{GREEN}{rgb}{0,1,0}
 \definecolor{BLUE}{rgb}{0,0,1}
 \definecolor{CYAN}{cmyk}{1,0,0,0}
 \definecolor{MAGENTA}{cmyk}{0,1,0,0}
 \definecolor{YELLOW}{cmyk}{0,0,1,0}
}

\@ifundefined{definecolor}
 {\usepackage{color}}{}
\usepackage{babel}
\hypersetup{colorlinks,citecolor=blue,filecolor=black,linkcolor=red,urlcolor=blue,pdftex}
\usepackage{colortbl}

\makeatother

\begin{document}

\title{Dark soliton in quasi-one-dimensional Bose-Einstein condensates
\\with a Gaussian trap}

\author{H. L. C. Couto}
\affiliation{Instituto de Física, Universidade Federal de Goiás, 74.001-970, Goiânia,
Goiás, Brazil}
\author{W. B. Cardoso}
\email{wesleybcardoso@ufg.br}
\affiliation{Instituto de Física, Universidade Federal de Goiás, 74.001-970, Goiânia,
Goiás, Brazil}

\begin{abstract}
In this paper we study dark solitons in quasi-one-dimensional Bose-Einstein
condensates (BECs) in presence of an anharmonic external potential.
The theoretical model is based on the Muñoz-Mateo \& Delgado (MMD)
equation that describes cigar-shaped BECs with repulsive interatomic
interactions. Since MMD equation presents a nonpolynomial form, the
soliton-sound recombination cannot display the same pattern presented
in the cubic model. We perform numerical simulations to compare both
cases.
\end{abstract}


\maketitle

\section{Introduction}

Solitons are localized structures that emerge from a perfect balance
between the dispersive and nonlinear effects in the system \cite{Kivshar03}.
In special, dark solitons are characterized by a depression in the
ambient density and a phase slip. This type of soliton is divided
into two classes: the \emph{black} ones, for which the minimum density
is zero, and the \emph{gray} ones, for which the dip in the density
is greater than zero. They were experimentally realized in nonlinear
optics \cite{KrokelPRL88}, shallow liquids \cite{DenardoPRL90},
magnetic films \cite{ChenPRL93}, ultracold atomic Bose-Einstein condensates
(BECs) \cite{BurgerPRL99,DenschlagSCI00,AndersonPRL01}, etc. In particular,
BECs with repulsive interatomic interaction are prone to generation
of dark solitons by various methods, e.g., by imprinting spatial phase
distribution \cite{DenschlagSCI00}, by inducing density defects in
BEC \cite{DuttonSCI01}, and by colision of two condensates \cite{ReinhardtJPB97,ScottJPB98}.

Differently from the case of attractive interatomic interaction, where
bright solitons emerge without necessity of an external potential,
due to the repulsive nature of atoms dark solitons requires an external
confining potential. Indeed, the harmonic potentials were used in
the experiments with BECs \cite{BurgerPRL99,DenschlagSCI00,AndersonPRL01}.
Also, for zero temperature in 1D regime, dark solitons are stable
and only solitons with zero velocity in the trap center do not move,
otherwise they oscillate along the trap axis \cite{BurgerPRL99,BuschPRL00}.
On the other hand, dark soliton propagating in an inhomogeneous condensate
has also been predicted to be unstable to the emission of sound waves
\cite{HuangPRA02,ParkerPRA10}. Recently, in Ref. \cite{ParkerPRA10}
was shown that such anharmonicities could break the soliton-sound
equilibrium and lead to the net decay of the soliton on a considerably
shorter time scale than other dissipation mechanisms.

In this paper we study numerically the effects of soliton-sound recombination
in presence of two different potentials: harmonic and Gaussian. Here,
differently from Refs. \cite{ParkerPRA10,ParkerPRL03,ProukakisPRL04},
we will use the Muñoz-Mateo \& Delgado (MMD) equation \cite{MMDPRA08,MateoPRA07},
which is an effective one-dimensional (1D) equation that governs the
axial dynamics of mean-field cigar-shaped condensates with repulsive
interatomic interactions, accounting accurately for the contribution
from the transverse degrees of freedom. To obtain this equation, in
Ref. \cite{MMDPRA08} the authors have used the standard adiabatic
approximation and an accurate analytical expression for the corresponding
local chemical potential in terms of the longitudinal density of the
condensate, expression which determine the form of the nonlinearity
\cite{MateoPRA07} (see next section).

The paper is organized as follows: in the next section we revisit
the theoretical model to obtain the 1D reduction of 3D Gross-Pitaevskii
equation according to Ref. \cite{MMDPRA08}; in sec. III, we show
the numerical procedure, considering the evolution of the ``soliton
part'' and the ``background part'', separately; the results
are shown in sec. IV for different patterns of potential; comments
and conclusion are displayed in sec. V.

\section{Theoretical Model}

The behavior of the wave function $\Psi(\mathbf{r},t)$ of the BEC
considering two-body interatomic interaction is well described by
the 3D Gross-Pitaevskii (GP) equation \cite{Pethick02} 
\begin{equation}
i\hbar\frac{\partial\Psi}{\partial t}=-\frac{\hbar^{2}}{2m}\nabla^{2}\Psi+V(\mathbf{r})\Psi+gN|\Psi|^{2}\Psi,\label{3d}
\end{equation}
 where $N$ is the number of atoms in the BEC, $g=4\pi\hbar^{2}a/m$
is the interaction strength, $a$ is the s-wave scattering length,
$m$ is the mass of the atomic specie, and $V(\mathbf{r})$ is the
external potential. In a cigar-shaped configuration, i.e., when the
frequency of transverse confinement is greater than the longitudinal
one ($\omega_{\perp}\gg\omega_{x}$), the wave function can be considered
with the form $\Psi(\mathbf{r},t)=\varphi(\mathbf{r_{\perp}};n(x,t))\phi(x,t),$
where $\varphi$ and $\phi$ are the transversal and longitudinal
wave functions, $\mathbf{r_{\perp}}=(y,z)$, and $n(x,t)$ is the
local density per unit length characterizing the axial configuration
$n\equiv N\int d^{2}\mathbf{r_{\perp}}|\Psi|^{2}=N|\phi|^{2}$, since
we have considered the wave function normalized to unity. In this
scenario, cf. Ref. \cite{MMDPRA08}, the transversal wave function
is adjusted instantaneously to the lowest-energy configuration compatible
with the axial configuration at each instant of time (adiabatic approximation).
Next, substituting the factorized wave function into the GPE, with
the potential given by $V(\mathbf{r})=V_{\perp}(r_{\perp})+V_{x}(x)$,
multiplying by $\varphi^{*}$ and integrating on the transverse coordinates
$\mathbf{r_{\perp}}$, one obtains 
\begin{equation}
i\hbar\frac{\partial\phi}{\partial t}=-\frac{\hbar^{2}}{2m}\frac{\partial^{2}\phi}{\partial x^{2}}+V_{x}\phi+\mu_{\perp}(n)\phi,\label{1d}
\end{equation}
 where we follow the definition of Ref. \cite{MMDPRA08} for the transversal
chemical potential with the form 
\begin{equation}
\mu_{\perp}\equiv\int d^{2}r_{\perp}\varphi^{*}\left[-\frac{\hbar^{2}}{2m}\nabla_{\perp}^{2}+V_{\perp}+ng|\varphi|^{2}\right]\varphi.\label{mu}
\end{equation}
 Then, one can find analytical solutions for $\mu_{\perp}$ in the
Eq. (\ref{mu}) for the two limit cases $an\ll1$ and $an\gg1$, such
that the dimensionless chemical potential takes the form $\overline{\mu}_{\perp}=1+2an$
(Gaussian approximation) and $\overline{\mu}_{\perp}=2\sqrt{an}$
(Thomas-Fermi approximation), respectively. According to Ref. \cite{MateoPRA07},
by using a suitable approximation scheme one can obtain $\overline{\mu}_{\perp}=\sqrt{1+4an}$,
that provide the ground-state properties of any mean-field scalar
Bose-Einstein condensate with short-range repulsive interatomic interactions,
confined in arbitrary cigar-shaped cylindrically symmetric harmonic
traps. In this case, we can conveniently rewrite the Eq. (\ref{1d})
in the dimensionless form 
\begin{equation}
i\frac{\partial\psi}{\partial t}=-\frac{1}{2}\frac{\partial^{2}\psi}{\partial x^{2}}+\overline{V_{x}}\psi+\lambda\sqrt{1+\sigma|\psi|^{2}}\psi,\label{MD1}
\end{equation}
 where we have considered $\phi=\psi/\sqrt{l_{x}}$, $t\rightarrow t/\omega_{x}$,
$x\rightarrow xl_{x}$, $V_{x}=\hbar\omega_{x}\overline{V_{x}}$,
$a=\overline{a}l_{x}$, and $\sigma=4\overline{a}N$, where $l_{x}=\sqrt{\hbar/m\omega_{x}}$
is the oscillator length in the axial direction and $\lambda=\omega_{\perp}/\omega_{x}\gg1$.

The goal of the present paper is to study dark solitons in the MMD
equation (\ref{MD1}) considering two different patterns of potentials
(harmonic and Gaussian) and comparing the results with the previous
studies of the cubic model \cite{ParkerPRL03,ProukakisPRL04,ParkerJPB04,MuryshevPRL02,BrazhnyiPRA03}.
We stress that the MMD equation (\ref{MD1}) is the effective 1D equation
that governs the axial dynamics of mean-field cigar-shaped condensates
with repulsive interatomic interactions, which incorporate more accurately
the contribution from the transverse degrees of freedom \cite{MMDPRA08}.
Also, in this regime it is more accurate when compared with the 1D
nonpolynomial Schrödinger equation obtained previously in Ref. \cite{SalasnichPRA04}\emph{.}

\section{Numerical procedure}

To solve Eq. (\ref{MD1}) we will use a numerical method based on
the split-step algorithm, which splits the time integration in two
parts: one containing the dispersive term of (\ref{MD1}) and the
other with the nondispersive terms, and solving the two parts separately.
More specifi{}cally, we will use the symmetric splitting method with
second-order accuracy in time \cite{Yang10}. Also, we use a Crank-Nicholson
algorithm to solve the dispersive term (for more details, see Ref.
\cite{MuruganandamCPC09}). Here, we have used the time and space
steps $\Delta t=0.001$ and $\Delta x=0.04$, respectively.

We will use an approximated input state to solve numerically the nonpolynomial
NLS equation (\ref{MD1}). The method used to get this input state
is similar to that studied in Ref. \cite{BrazhnyiPRA03} for the cubic
nonlinear Schrödinger equation. To this end, we will consider the
evolution of the ``soliton part'' and the ``background part'',
separately. Then, we will employ the following ansatz 
\begin{equation}
\psi=\rho(x,t)\Phi(x,t),\label{ansatz}
\end{equation}
 where $\rho(x,t)$ is the inhomogeneous background and $\Phi(x,t)$
is the dark soliton solution (homogeneous background). Next, replacing
the Eq. (\ref{ansatz}) into (\ref{MD1}) one gets 
\begin{eqnarray}
i\rho_{t}\Phi+i\rho\Phi_{t} & = & -\frac{1}{2}\rho_{xx}\Phi-\rho_{x}\Phi_{x}-\frac{1}{2}\rho\Phi_{xx}\nonumber \\
 & + & \overline{V_{x}}\rho\Phi+\lambda\sqrt{1+\sigma\rho^{2}|\Phi|^{2}}\rho\Phi.\label{s1}
\end{eqnarray}
 We hope $\rho$ to be a nodeless background solution that satisfy
the following equation 
\begin{equation}
i\rho_{t}=-\frac{1}{2}\rho_{xx}+\overline{V_{x}}\rho+\lambda\sqrt{1+\sigma\rho^{2}|\Phi_{\infty}|^{2}}\rho,\label{background}
\end{equation}
 where $|\Phi_{\infty}|$ is the absolute value of the function $\Phi$
with $x\rightarrow\pm\infty$. So, defining $\rho\equiv f(x,t)/|\Phi_{\infty}|$,
we can rewrite the Eq. (\ref{background}) without the dependence
of $|\Phi_{\infty}|$, given by 
\begin{equation}
if_{t}=-\frac{1}{2}f_{xx}+\overline{V_{x}}f+\lambda\sqrt{1+\sigma f^{2}}f.\label{rescaled_back}
\end{equation}
Eq. (\ref{rescaled_back}) can be solved by using the imaginary time
propagation method in which a stable (nodeless) solution with lower
energy emerges (background).

\begin{figure}[tb]
\centering \includegraphics[width=0.8\columnwidth]{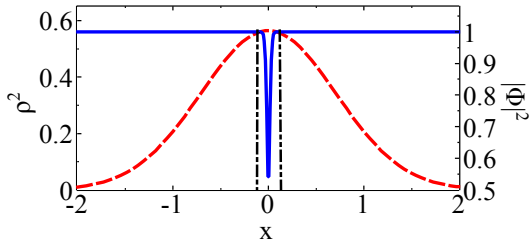}

\caption{(Color online) Schematic representation of the regions of variation
for each function. The ``background solution'' and the ``soliton
solution'' are represented by dashed (red) and solid (blue) lines,
respectively. Note that the left- and right-vertical curves in dash-dot
(black) line delimit the region of variation of the two functions
(approximately). While $\rho^{2}(x)$ varies on the outside, $|\Phi(x,0)|^{2}$
varies inside.}

\label{F1} 
\end{figure}

Since the Eq. (\ref{background}) is satisfied, Eq. (\ref{s1}) assumes
the form 
\begin{eqnarray*}
i\Phi_{t} & = & -\frac{1}{2}\Phi_{xx}-(\ln\rho)_{x}\Phi_{x}+\lambda\sqrt{1+\sigma\rho^{2}|\Phi|^{2}}\Phi\\
 & - & \lambda\sqrt{1+\sigma\rho^{2}|\Phi_{\infty}|^{2}}\Phi.
\end{eqnarray*}
 The above equation can be conveniently rewrited such that 
\begin{eqnarray}
i\Phi_{t} & = & -\frac{1}{2}\Phi_{xx}+\lambda\sqrt{1+\sigma\rho_{0}^{2}|\Phi|^{2}}\Phi\nonumber \\
 & - & \lambda\sqrt{1+\sigma\rho_{0}^{2}|\Phi_{\infty}|^{2}}\Phi+R(\rho,\Phi),\label{homogeneous}
\end{eqnarray}
 where 
\begin{eqnarray}
R(\rho,\Phi) & = & -(\ln\rho)_{x}\Phi_{x}+\lambda\sqrt{1+\sigma\rho^{2}|\Phi|^{2}}\Phi\nonumber \\
 & - & \lambda\sqrt{1+\sigma\rho^{2}|\Phi_{\infty}|^{2}}\Phi+\lambda\sqrt{1+\sigma\rho_{0}^{2}|\Phi_{\infty}|^{2}}\Phi\nonumber \\
 & - & \lambda\sqrt{1+\sigma\rho_{0}^{2}|\Phi|^{2}}\Phi.\label{R}
\end{eqnarray}
 Note that the Eq. (\ref{homogeneous}) takes the form of a homogeneous
nonlinear equation when $R$ vanishes. Also, since the background
$\rho$ does not change in the region of strong variation for $\Phi$,
and \emph{vice-versa}, we can consider 
\begin{equation}
(\ln\rho)_{x}\Phi_{x}\simeq0.\label{s2}
\end{equation}
 Fig. \ref{F1} shows a schematic representation of the regions of
variation for each function.

The result of Eq. (\ref{s2}) becomes exact for the limit cases $x\rightarrow\pm\infty$
($\Phi_{\infty}$ constant) and $x\rightarrow0$ (since the trap provides
a background centered in $x=0$). Also, the Eq. (\ref{R}) leads to
a null result in these limit cases.

Now, defining $\mu=\lambda\sqrt{1+\sigma\rho_{0}^{2}|\Phi_{\infty}|^{2}}$
and assuming $R(\rho,\Phi)\simeq0$ in (\ref{homogeneous}), we will
have 
\begin{equation}
i\Phi_{t}+\mu\Phi=-\frac{1}{2}\Phi_{xx}+\lambda\sqrt{1+\sigma\rho_{0}^{2}|\Phi|^{2}}\Phi.\label{homogeneous_2}
\end{equation}
 Next, using the rescaling $\Phi\equiv\widetilde{\Phi}/\rho_{0}$
one gets 
\begin{equation}
i\widetilde{\Phi}_{t}+\mu\widetilde{\Phi}=-\frac{1}{2}\widetilde{\Phi}_{xx}+\lambda\sqrt{1+\sigma|\widetilde{\Phi}|^{2}}\widetilde{\Phi}.\label{homogeneous_3}
\end{equation}
 Following, to solve the Eq. (\ref{homogeneous_3}) we will use the
ansatz 
\begin{equation}
\widetilde{\Phi}=A(\zeta)e^{i\eta(\zeta)},\label{sol_homogeneous}
\end{equation}
 where $\zeta=x-vt$ with $v$ being the initial soliton velocity,
and $A$ and $\eta$ are real functions. Inserting (\ref{sol_homogeneous})
in (\ref{homogeneous_3}) we obtain the imaginary part satisfying
\begin{equation}
\eta_{\zeta}=v\left(1-\frac{A_{\infty}^{2}}{A^{2}}\right),\label{phase}
\end{equation}
 where $A_{\infty}$ is the value of $A(\zeta=\infty)$ and we have
used $\lim_{\zeta\rightarrow\pm\infty}(\eta_{\zeta})=0$; the real
part, considering the result of (\ref{phase}) evolves to 
\begin{equation}
A_{\zeta\zeta}=-(v^{2}+2\mu)A+v^{2}\frac{A_{\infty}^{4}}{A^{3}}+2\lambda\sqrt{1+\sigma A^{2}}A.\label{Azz}
\end{equation}
 The above equation can be reduced to a first order differential equation
given by 
\begin{equation}
A_{\zeta}=\pm\sqrt{2U+c_{1}},\label{first_order}
\end{equation}
 where $U=-\frac{1}{2}(v^{2}+2\mu)A^{2}-(v^{2}A_{\infty}^{4}/2A^{2})+\frac{2\lambda}{3\sigma}(1+\sigma A^{2})^{3/2}$
and $c_{1}=2(v^{2}+\mu)A_{\infty}^{2}-\frac{4\lambda}{3\sigma}(1+\sigma A_{\infty}^{2})^{3/2}$,
since we have consider $\lim_{\zeta\rightarrow\infty}(A_{\zeta})=0$.
Now, considering $\lim_{\zeta\rightarrow\infty}(A_{\zeta\zeta})=0$
in Eq. (\ref{Azz}) one obtains 
\begin{equation}
A_{\infty}^{2}=\frac{1}{\sigma}\left(\frac{\mu^{2}}{\lambda^{2}}-1\right).\label{A00}
\end{equation}
 Note that for $A_{\infty}$ to be real one needs $\mu>\lambda$ ($\lambda$
is a positive constant as we had previously defined and $\sigma>0$).
Also, the pattern of $U$ in (\ref{first_order}) presents a local
minimum corresponding to the value of $A_{\infty}$, such that $\left.U_{AA}\right|_{A_{\infty}}>0$
and consequently $\mu>v^{2}+\sqrt{v^{4}+\lambda^{2}}$. So, one gets
a limit value for the chemical potential in function of the frequencies
ratio $\lambda$ and the soliton velocity $v$.

Following, taken $\lim_{\zeta\rightarrow0}\left(A^{2}\right)_{\zeta}=0$
in Eq. (\ref{first_order}) we will obtain a quintic order equation
in $A_{0}$, such that, by using the rescale $1+\sigma A_{0}^{2}\equiv\gamma^{2}$
one gets 
\begin{equation}
a\gamma^{5}+b\gamma^{4}+c\gamma^{3}+d\gamma^{2}+e=0,\label{quintic}
\end{equation}
 where $a=4\lambda^{5}$, $b=-3\lambda^{4}(v^{2}+2\mu)$, $c=-a$,
$d=2\mu\lambda^{2}(3v^{2}\mu+3\lambda^{2}+\mu^{2})$, and $e=-\mu^{3}(3v^{2}\mu+2\lambda^{2})$.
The Eq. (\ref{quintic}) admit two solutions $\gamma=\gamma_{\infty}=\mu/\lambda$.
In this case one can reduct the Eq. (\ref{quintic}) for a third order
equation given by $a^{\prime}\gamma^{3}+b^{\prime}\gamma^{2}+c^{\prime}\gamma+d^{\prime}=0$,
with $a^{\prime}=4\lambda^{3}$, $b^{\prime}=\lambda^{2}(2\mu-3v^{2})$,
$c^{\prime}=-2\lambda(3\mu v^{2}+2\lambda^{2})$, and $d^{\prime}=-\mu(3\mu v^{2}+2\lambda^{2})$.
We have used the Cardano's method to solve analytically this cubic
equation. Indeed, the cubic equation has always one real positive
root. We have used this root to get the value of $A_{0}$ and start
a numerical method ($4^{th}$ order Runge-Kutta) to get the profiles
of $A(\zeta)$ and $\eta(\zeta)$. So, Eq. (\ref{sol_homogeneous})
gives us the initial profile ($t=0$) of the homogeneous part of the
ansatz (\ref{ansatz}). Note that the Eq. (\ref{ansatz}) takes the
following form $\psi=f\widetilde{\Phi}/\rho_{0}|\Phi_{\infty}|$.
Then, using our definition of $\mu$ we will have $\rho_{0}|\Phi_{\infty}|=\sqrt{(\mu/\lambda)^{2}-1}/\sqrt{\sigma}$
that is the transformation factor for the rescaled solutions. The
last step consists in the use of $\psi(x,0)$ as the initial profile
of the split-step algorithm to solve the Eq. (\ref{MD1}).

\section{Results}

Next we will show the numerical results considering two different
patterns of potential. In order to compare the results with the cubic
case (Ref. \cite{ParkerPRA10}), we will use here the expansion in
first order 
\begin{equation}
\lambda\sqrt{1+\sigma|\psi|^{2}}\simeq\lambda+(\lambda\sigma/2)|\psi|^{2}.\label{approx}
\end{equation}
 Note that in the cubic approximation the chemical potential is rescaled
by $\lambda$, i.e., $\mu_{C}\equiv\mu-\lambda$, where $\mu_{C}$
($\mu$) is the chemical potential for the cubic (nonpolynomial) case.
Also, the cubic nonlinearity takes the corresponding relationship
$\sigma_{C}=\lambda\sigma/2$. In the general case the two equalities
above are not valid simultaneously due to the approximation (\ref{approx}).
So, we will establish the equality between the nonlinearities, leaving
aside the relationship between the chemical potentials to investigate
in the next subsection the influence of a harmonic trap. We will name
the cubic equation as 1D GP equation from now on.

\subsection{Harmonic trap}

The standard case consists on the quadratic potential that confines
the BEC 
\begin{equation}
\overline{V_{x}}=x^{2}/2.
\end{equation}
 In this case a dark soliton (obtained following Eq. (\ref{ansatz}))
with an initial velocity at center of the BEC given by $v_{0}=0.5v_{l}$,
where $v_{l}=\sqrt{\mu}$ and $v_{l}=\sqrt{\mu^{2}-\lambda^{2}}/\sqrt{2\mu}$
for the cubic and nonpolynomial case, respectively, evolves such that
the velocity of dark soliton is reduced and the depth of the dark
soliton is increased (reducing its velocity) until it touch the zero
density. At this point, the velocity of the soliton changes its direction
allowing an oscillatory pattern (like a particle in a harmonic oscillator).
However, due to the soliton acceleration it emits a shock wave (sound
wave). In the present case, the recombination soliton-sound maintains
a stable solution.

\begin{figure}[tb]
\centering \includegraphics[width=0.75\columnwidth]{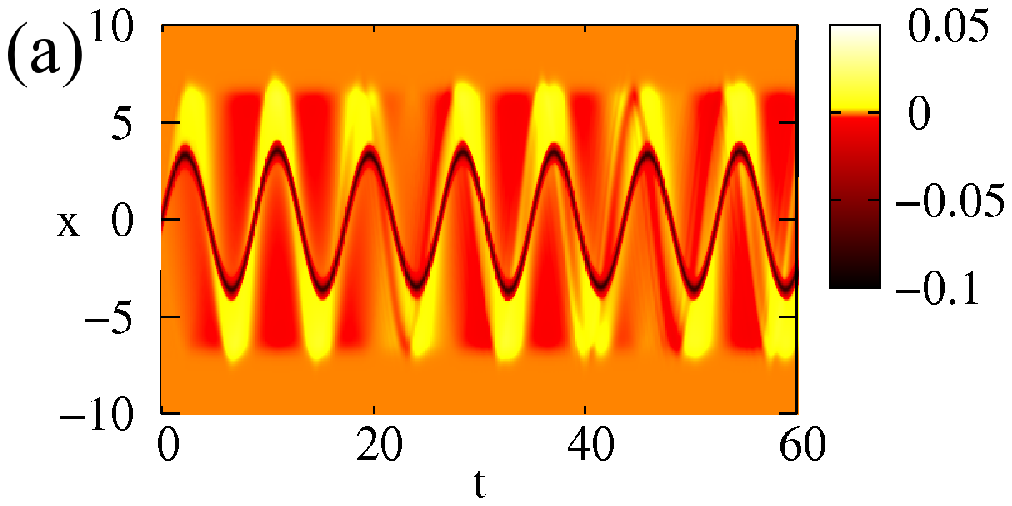} \includegraphics[width=0.9\columnwidth]{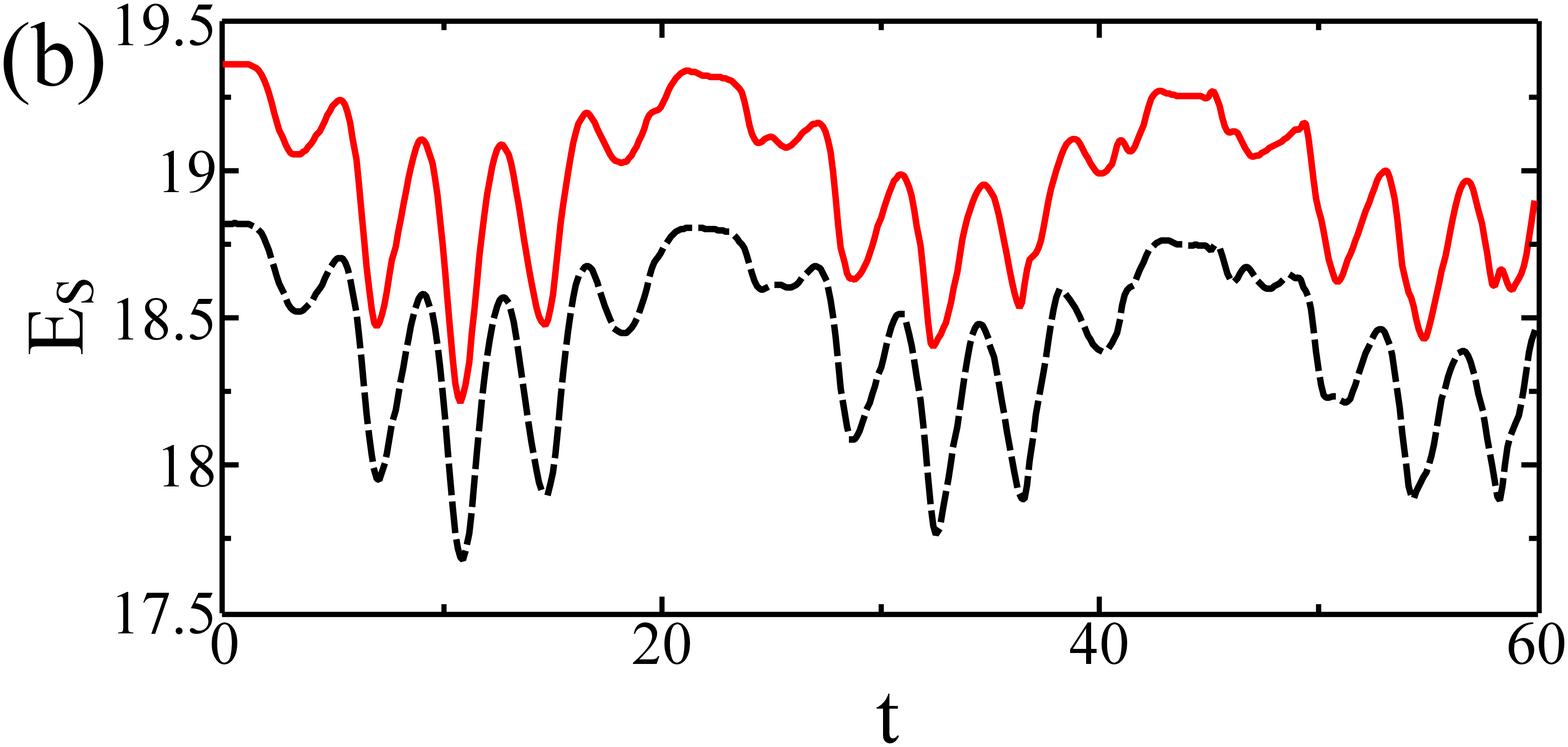}

\caption{(Color online) Dark soliton dynamics in a harmonic trap. The renormalized
density $|\psi|^{2}-f^{2}$ for the 1D GP equation are displayed in
(a). The behavior of the dark soliton in the MMD equation is similar
to the cubic case. (b) Soliton energy for the cubic in (red) solid
line and nonpolynomial (black) dashed line, both in in units of $\hbar\omega_{z}\rho_{0}^{2}\xi$
(the dashed gray bars assist us to the visualization of the similar
patterns). Here we have considered $v_{0}=0.5v_{l}$, where $v_{l}=\sqrt{\mu}$
and $v_{l}=\sqrt{\mu^{2}-\lambda^{2}}/\sqrt{2\mu}$ for the cubic
and nonpolynomial case, respectively; $\sigma_{C}=200$, $\sigma=2$
and $\lambda=200$, providing $\mu_{C}\simeq22.42$ (which is close
to that used in Ref.\cite{ParkerPRA10}) and $\mu-\lambda\simeq21.76$.}

\label{x2_01} 
\end{figure}

Fig. \ref{x2_01}(a) shows the renormalized density $|\psi|^{2}-f^{2}$
for the 1D GP equation as a function of time (similar results were
verified for MMD). Sound waves are in light blur while the soliton
position is in the dark trail. In Fig. \ref{x2_01}(b) we display
the temporal evolution of the soliton energy (see Appendix). The soliton
position (as well as the mean position of the BEC, defined by $\overline{x}=\int_{-\infty}^{\infty}x|\psi|^{2}dx$)
is coincident for both cases considering the parameters $\sigma_{C}=200$,
$\sigma=2$ and $\lambda=200$. Note that the approximation (\ref{approx})
is more accurate for small values of $\sigma|\psi|^{2}$. So, the
smaller $\sigma|\psi|^{2}$ is, since the relation $\sigma_{C}=\lambda\sigma/2$
is satisfied, better is the match for all calculated quantities comparing
the results of the evolution in both models. This was confirmed in
our numerical simulations. However, we stress that even considering
the above relation between the nonlinearities, we need $\lambda\gg1$
to be valid the 1D approximation.

In contrast with the above result, when considering $\sigma_{C}=2000$,
$\sigma=100$ and $\lambda=40$, satisfying $\sigma_{C}=\lambda\sigma/2$,
we have obtained a discrepant set of quantities. For example, the
chemical potential is obtained to be $\mu_{C}=104$.00 and $\mu-\lambda=72.35$.
Also, we have used the power spectrum of some functions to obtain
the principal frequency contributions in these two cases. As expected,
the oscillation frequency of the center of mass of the BEC is $\omega_{x}=1$
for the two cases. However, the oscillation frequencies for the solitonic
position are $\omega_{C}\simeq0.708$ and $\omega_{MMD}\simeq0.72$.
So, for the cubic case the relation $\omega_{x}/\omega_{C}=\sqrt{2}$
is satisfied while in the nonpolynomial case there is $2\%$ of error.
This is also verified by using the energy oscillation.

\begin{figure*}[tb]
\centering \includegraphics[width=0.3\textwidth]{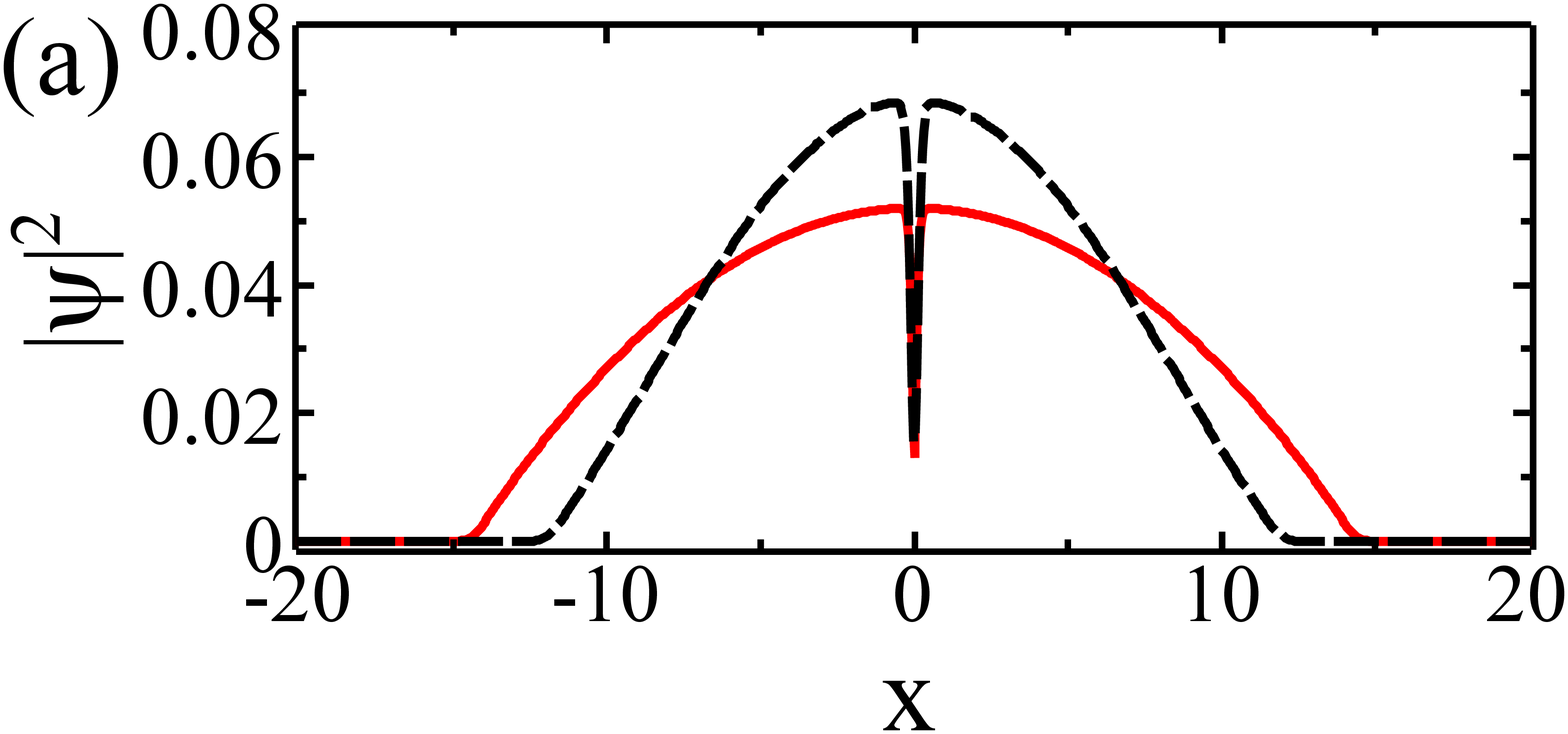} \includegraphics[width=0.3\textwidth]{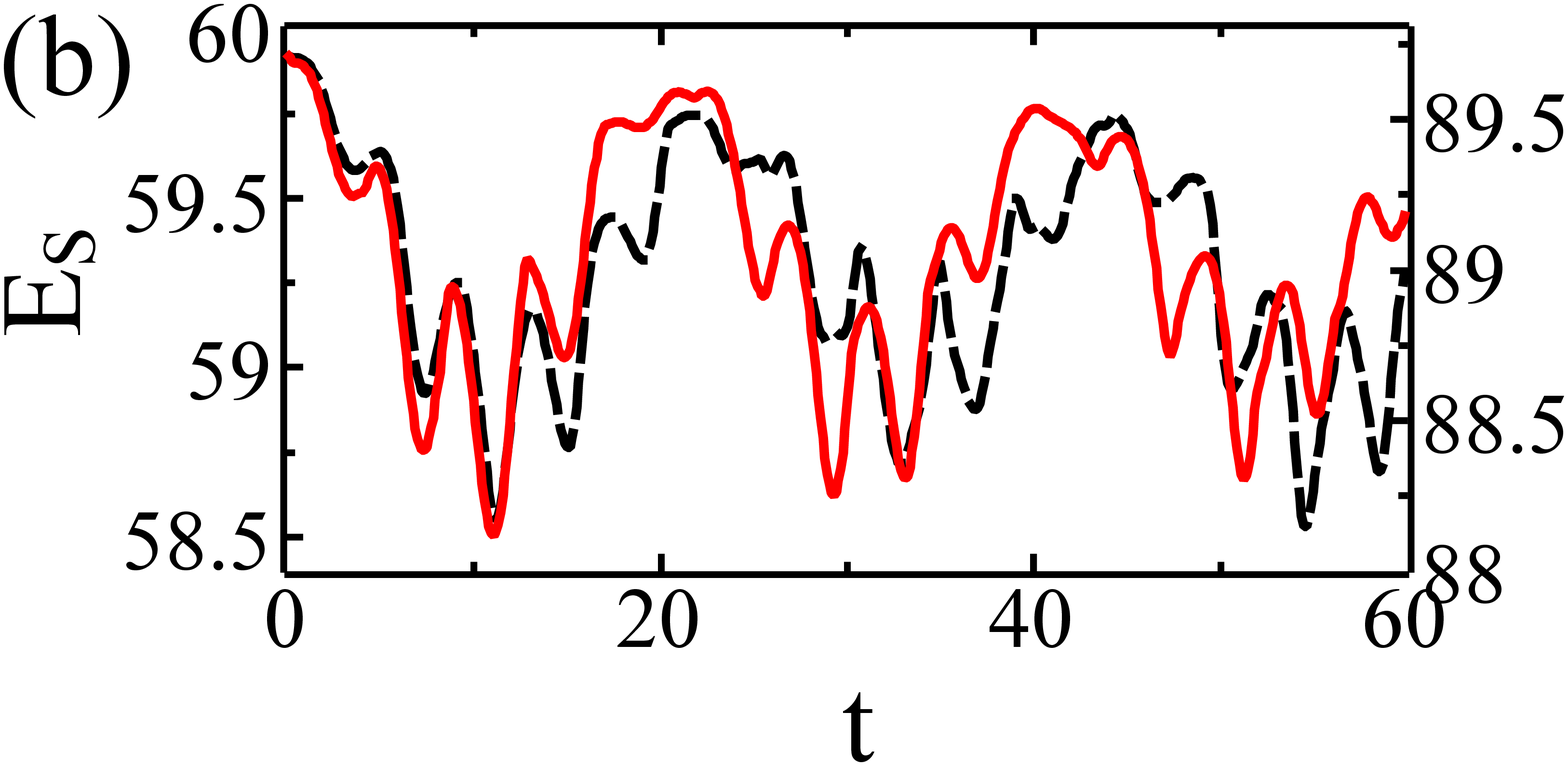}
\includegraphics[width=0.3\textwidth]{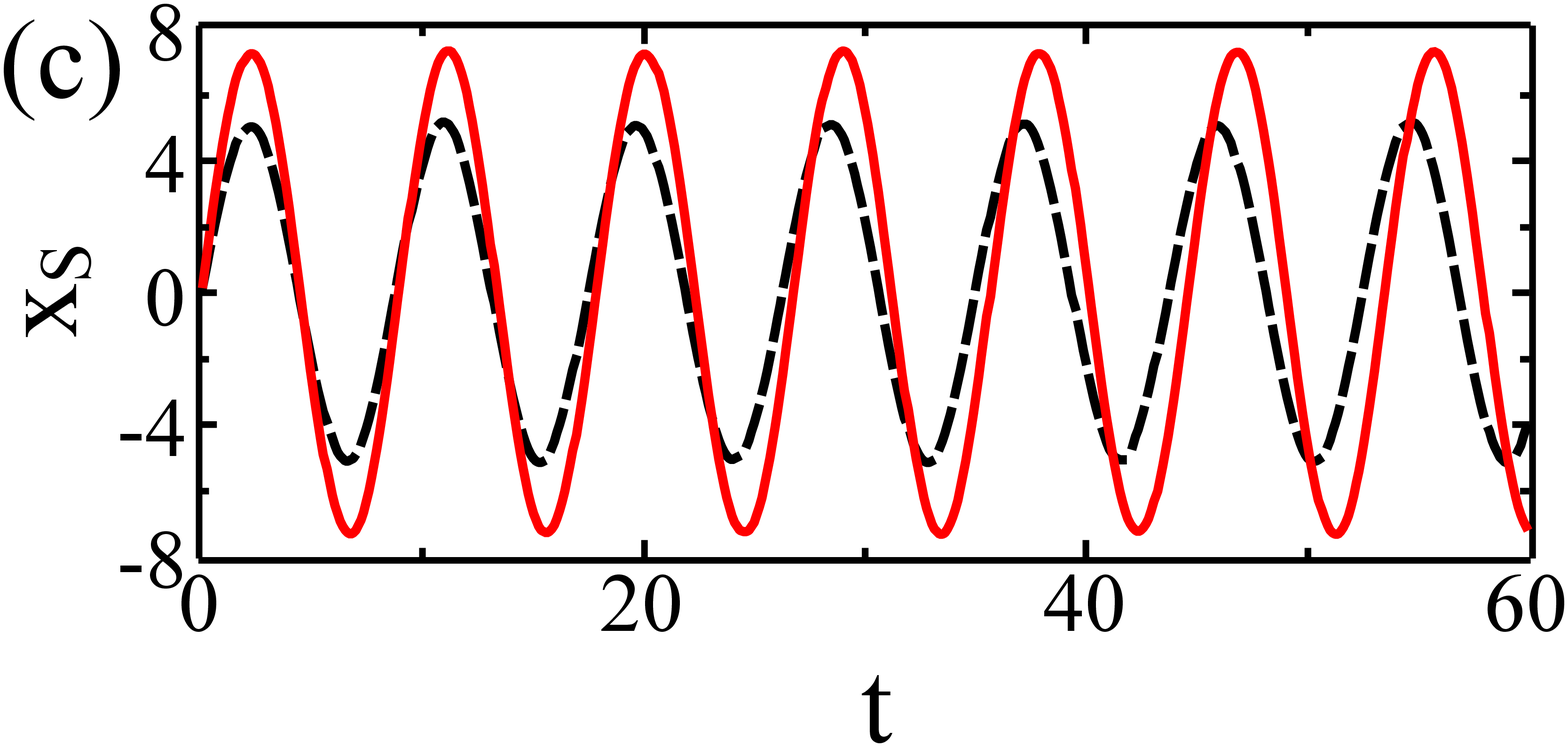} \caption{(Color online) Dark soliton dynamics in presence of an harmonic trap.
(a) Input profiles. (b) Comparison between the energy values as function
of time. Soliton energy values considering the MMD (GP) equation are
displayed in left (right) axis. (c) Soliton position. The solid (red)
lines represent the results for the GP equation while the results
for the MMD equation is presented in dashed (black) lines. We have
used $\sigma_{C}=2000$, $\sigma=100$ and $\lambda=40$.}

\label{x2_02} 
\end{figure*}

Fig. \ref{x2_02}(a) shows the input profiles for the two equations.
Note that the solitonic profile seems similar but the background is
more localized for the MMD equation. This evident contrast is verified
in the energy scales displayed in Fig. \ref{x2_02}(b) (left and right
axis), as well as the difference between its oscillatory behaviors.
The soliton positions for the two cases are shown in Fig. \ref{x2_02}(c).

\subsection{Gaussian trap}

Here, we will consider a Gaussian trap of the form 
\begin{equation}
\overline{V_{x}}=V_{0}\left[1-\exp\left(-x^{2}/2V_{0}\right)\right],\label{gauss_trap}
\end{equation}
 where $V_{0}$ is the depth of the trap. Firstly we want to know
the influence of cutoff value $V_{0}$ to the soliton dynamics. To
this end, we will fix a value for the nonlinearity in the 1D GP and
MMD equations (namely, $\sigma_{C}=\lambda\sigma/2=1200$). Then,
in this case the relation $\mu_{C}=\mu-\lambda$ will not be satisfied.

Since, by Thomas-Fermi approximation the BEC is concentrated in the
region $\overline{V_{x}}<\mu-\lambda$, when $V_{0}<\mu-\lambda$
in (\ref{gauss_trap}), sound waves can scape of the trap. On the
other hand, when $V_{0}\gg\mu-\lambda$, the sound waves are trapped
and the potential is approximately harmonic in the BEC region.

\begin{figure*}[tp]
\centering \includegraphics[width=0.3\textwidth]{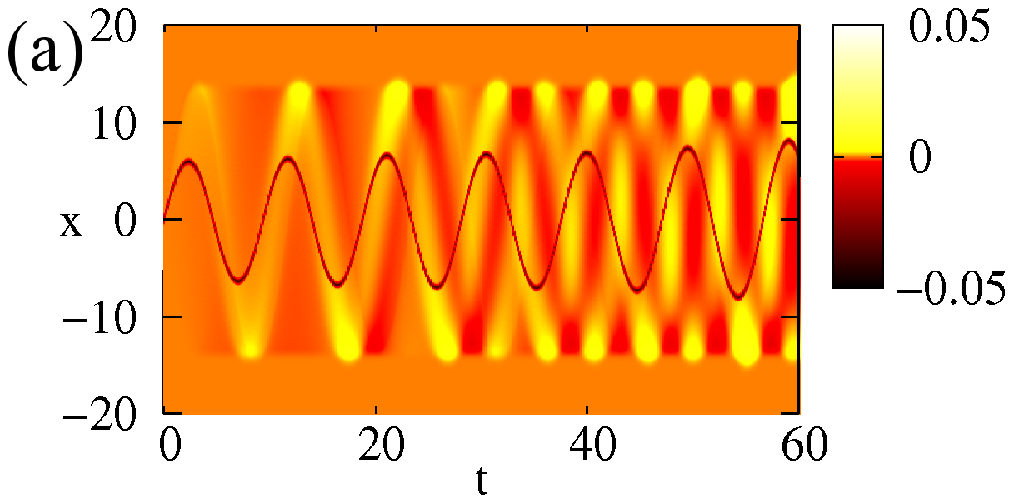} \includegraphics[width=0.3\textwidth]{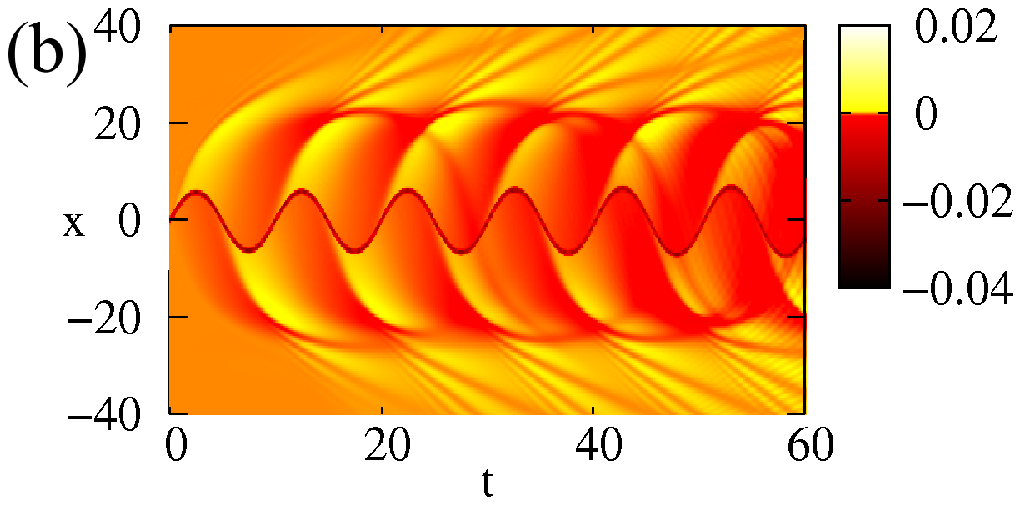}
\includegraphics[width=0.3\textwidth]{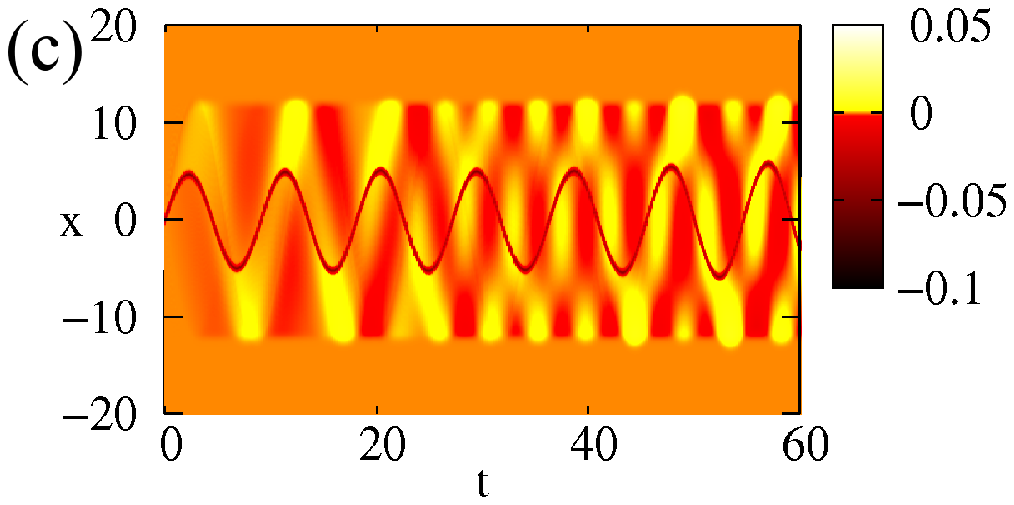} \includegraphics[width=0.3\textwidth]{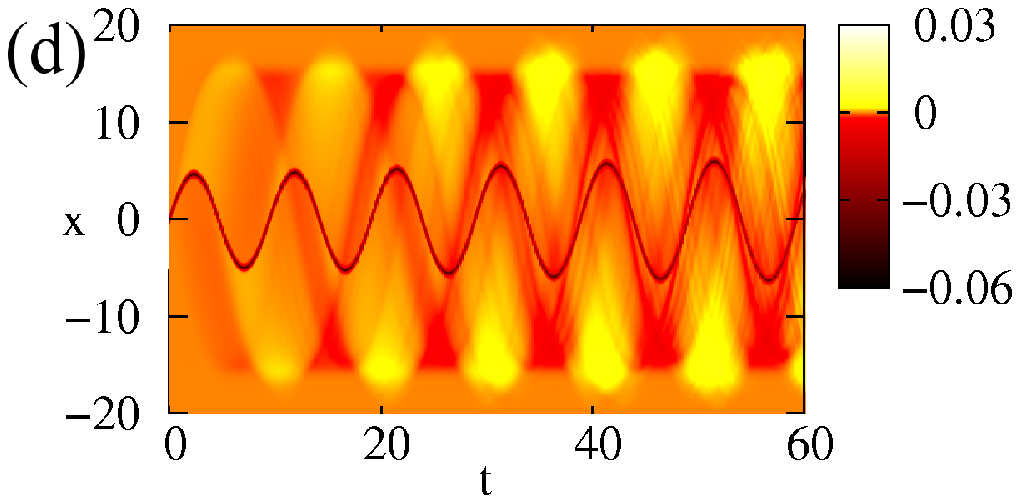}
\includegraphics[width=0.3\textwidth]{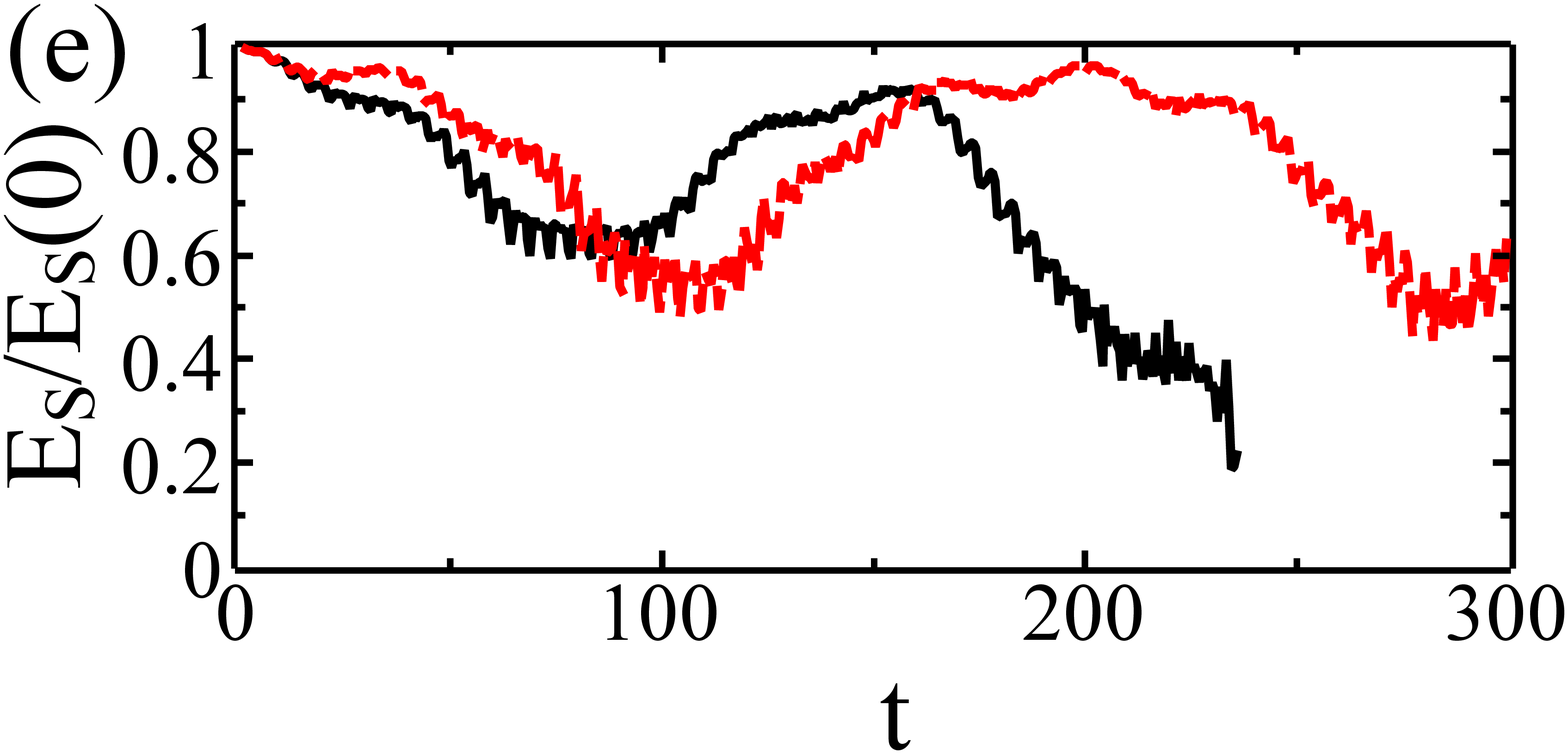} \includegraphics[width=0.3\textwidth]{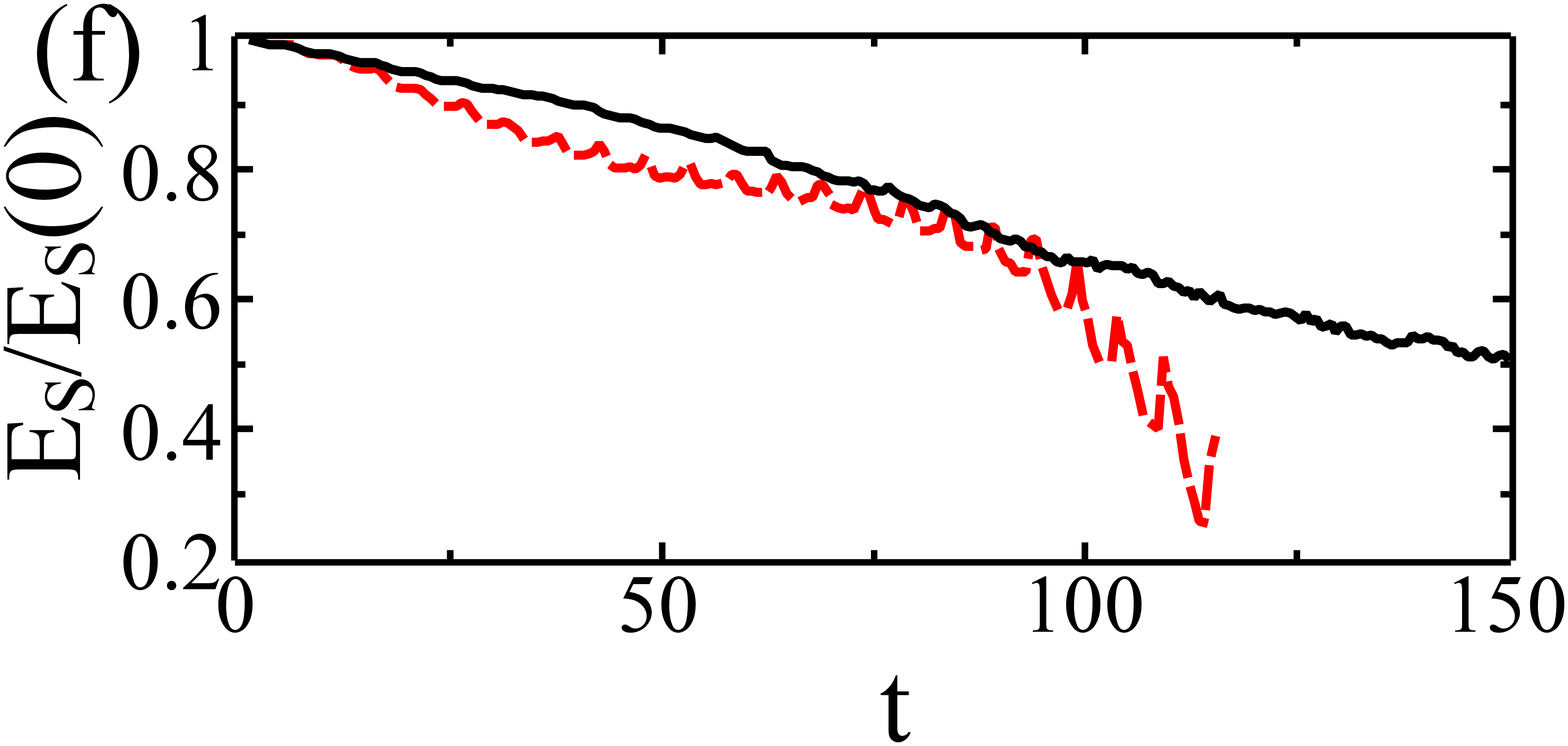}

\caption{(Color online) Renormalized density profile of dark soliton ($|\psi|^{2}-f^{2}$)
for the cubic (a) and (b) and nonpolynomial nonlinearity (c) and (d).
We have used $\sigma_{C}=\lambda\sigma/2=1200$ ($\sigma=40$ and
$\lambda=60$) with $V_{0}=2\mu_{C}$ in (a) and (c) and $V_{0}=\mu_{C}$
in (b) and (d). Evolution of the renormalized soliton energy $E_{S}/E_{s}(t=0)$
in a Gaussian trap for (e) $V_{0}=2\mu_{C}$ and (f) $V_{0}=\mu_{C}$.
The results of the 1D GP and MMD equations are displayed in solid
(black) and dashed (red) lines, respectively. }

\label{gauss_01} 
\end{figure*}

These results are shown in Figs. \ref{gauss_01}(a) for $V_{0}=2\mu_{C}$
(trapped) and \ref{gauss_01}(b) for $V_{0}=\mu_{C}$ (sound escapes),
considering the 1D GP equation. In Figs. \ref{gauss_01}(c) and \ref{gauss_01}(d)
we display the rescaled soliton profile for the MMD equation considering
$V_{0}=2\mu_{C}$ and $V_{0}=\mu_{C}$, respectively. Note that for
$V_{0}=\mu_{C}$ ($\mu_{C}\simeq61.19$) the sound escapes in the
1D GP equation but it does not escapes in the MMD equation. This is
evident once we have abdicated to the equality for the chemical potential
and its correct value in the MMD equation is to be $\mu-\lambda\simeq51.29$
when we set $\sigma=40$ and $\lambda=60$, satisfying the relation
$\sigma_{C}=\lambda\sigma/2$.

The temporal evolution of the rescaled soliton energies are shown
in Figs. \ref{gauss_01}(e) and \ref{gauss_01}(f) for $V_{0}=2\mu_{C}$
and $V_{0}=\mu_{C}$, respectively. The results for the 1D GP equation
are displayed in solid (black) lines while dashed (red) lines represent
the MMD equation. It is clear by Fig. \ref{gauss_01}(f) the dissipative
behavior when considering the cubic nonlinearity in opposition to
the trapped form when the nonlinearity is nonpolynomial. For the latter,
the soliton-sound recombination destroys the soliton faster than the
dissipative case given by 1D GP equation. Also, in Fig. \ref{gauss_01}(e)
one can see that the soliton lifetime is different in both cases,
i.e., $t=235.8$ ($t\simeq6.4$s) for the cubic case and $t>300$
($t\gtrsim8$s) for the nonpolynomial case.

\begin{figure}[tb]
\centering

\includegraphics[width=0.9\columnwidth]{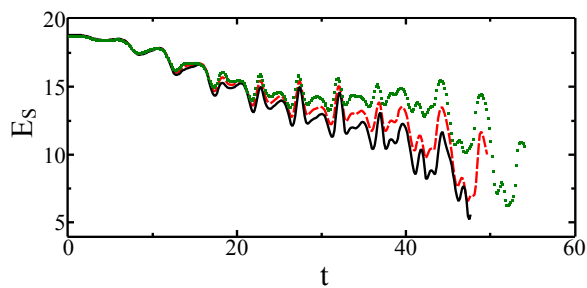}

\caption{(Color online) Temporal evolution of the soliton energy $E_{S}$ for
a Gaussian trap with cutoff $V_{0}=2(\mu-\lambda)$. Solid (black)
line corresponds to $\lambda=200$, in dashed (red) line $\lambda=100$,
and in doted (green) line $\lambda=50$. For all cases we have used
$\mu-\lambda=21.76$.}

\label{F5}
\end{figure}

To verify the influence of $\lambda$ in the soliton-sound recombination
we display in Fig. \ref{F5} the temporal evolution of the soliton
energy $E_{S}$ for the nonpolynomial case, considering $\lambda=200$
in solid (black) line, $\lambda=100$ in dashed (red) line, and $\lambda=50$
in doted (green) line. We have used the gaussian depth $V_{0}=2(\mu-\lambda)$,
with $\mu-\lambda=21.76$. Note that decreasing the value of $\lambda$
the lifetime of the soliton is increased. When $\lambda=200$, $\lambda=100$,
and $\lambda=50$ the corresponding soliton lifetimes are $t\simeq47.7\omega_{x}^{-1}$,
$t\simeq49.6\omega_{x}^{-1}$, and $t\simeq54.1\omega_{x}^{-1}$,
respectively. This accounts a difference of $\sim13\%$ comparing
the lifetime of the first and last cases. 

We now attempt to the experimental parameters obtained in Ref. \cite{BeckerNP08}.
In \cite{BeckerNP08} $\omega_{x}=2\pi\times5.9\mathrm{Hz}$ and $\omega_{\perp}\sim2\pi\times109\mathrm{Hz}$,
which leads to $\lambda\simeq18.5$. The chemical potential is $\mu_{C}/k_{B}\simeq20$
nK, where $k_{B}$ is the Boltzmann constant. In this case, one can
estimate $\mu_{C}\simeq70.6$ and consequently $\sigma_{C}\simeq1136.02$
by using the relation $\mu_{C}=\sigma_{C}n_{0}$ with $n_{0}$ obtained
through the profile obtained by propagation in imaginary time of the
1D GP equation with $V_{0}\gg\mu_{C}$ ($V_{0}=10\mu_{C}$). Following,
using $\sigma_{C}$ and $\lambda$ given above we obtain numerically
$\mu\simeq63.4$ (for $V_{0}=10(\mu-\lambda)$).

Next, by using the value of $\mu_{C}$ given above we estimate using
the profile obtained via the imaginary time propagation for the cubic
equation ($V_{0}\gg\mu_{C}$), a density peak for the background and
consequently the non-linearity intensity given by $\rho_{0}^{2}\simeq6.2\times10^{-2}$
and $\sigma_{C}\simeq1118.5$, respectively. Also, through $\mu$
above, we obtain $\rho_{0}^{2}\simeq8.3\times10^{-2}$ and $\sigma\simeq163.2$.

\section{Conclusion}

In conclusion, we have studied the soliton-sound interaction in the
MMD equation, which is an effective 1D equation governing the axial
dynamics of a cigar-shaped BEC with repulsive interatomic interactions,
accounting accurately for the contribution from the transverse degrees
of freedom. A significant differences has been observed when comparing
the soliton dynamics in MMD and 1D GP equation. In particular, increasing
the strength of the repulsive interatomic interaction the divergence
between the results appears naturally. Also, the soliton-sound recombination
presents an important hole in the lifetime of dark solitons, as shown
in the literature \cite{ParkerJPB04,ParkerPRA10,ParkerPRL03}. When
the perfect recombination does not occurs, for example in anharmonic
traps like that presented here, the soliton can scape from the trap
or simply to decay. This is in agreement with the results obtained
in the present paper. We believe that this can motivate further investigations
of soliton- or vortex-sound interactions in realistic systems with
more dimensions.

\section*{Acknowledgments}

We thank the CNPq, CAPES, and Instituto Nacional de Ciência e Tecnologia
- Informação Quântica (INCT-IQ), Brazilian agencies, for the partial
support.

\section*{Appendix: Dark soliton energy}

In presence of a confining potential the energy of the solution is
a finite constant. However, we want to compute only the contribution
of the dark soliton energy. To this end, we will use the renormalized
energy density, given by \cite{Kivshar03} 
\begin{equation}
\epsilon(\psi)=\frac{1}{2}|\psi_{x}|^{2}+\overline{V}|\psi|^{2}+\int_{f_{0}^{2}}^{|\psi|^{2}}[F(I)-F(f_{0}^{2})]dI,\label{e_density}
\end{equation}
where $F(I)=\sigma_{C}I$ and $F(I)=\lambda\sqrt{1+\sigma I}$ for
the cubic and nonpolynomial NLS equation, respectively. In the case
of cubic nonlinearity the Eq. (\ref{e_density}) reduces to a similar
form of Eq. (A2) of Ref. \cite{ParkerPRA10}. Next we will use the
following definition for the soliton energy 
\begin{equation}
E_{s}=\int_{x_{s}-x_{int}}^{x_{s}+x_{int}}\epsilon(\psi)dx-\int_{x_{s}-x_{int}}^{x_{s}+x_{int}}\epsilon(f)dx,
\end{equation}
where $f$ is the time-independent background density in the absence
of the soliton (i.e., the solution from the imaginary time propagation
of the Eq.(\ref{rescaled_back})) and $f_{0}$ denotes the peak condensate
density at the center of the trap for purely harmonic confinement.
Here, $x_{s}$ is the soliton position and $x_{int}$ is the domain
of integration around the soliton position. To find $x_{int}$ we
have varied its value until $E_{s}$ does not change anymore (we have
stopped the variation of $x_{int}$ with a difference of energy of
the order of $10^{-14}$). This value is compared to that obtained
in Ref. \cite{ParkerPRA10}.

\end{document}